\documentclass[
    ,final            % use final for the camera ready runs 
    ,numberedheadings % uncomment this option for numbered sections  
    ,cmfonts          % add further options here if necessary  
  ]  
  {aipproc}  
  
\layoutstyle{6x9}  
  
\begin{document}
\newcommand{\BA}{\begin{eqnarray}}
\newcommand{\EA}{\end{eqnarray}}  
\newcommand{\BE}{\begin{equation}}
\newcommand{\EE}{\end{equation}}
  
\title{QCD -- NLC}  
  
  
\author{H.J. Pirner\thanks{Supported by the EU-Program Computational  
  Hadron Physics}}{  
  address={University of Heidelberg, Germany \\  
E-mail: pir@tphys.uni-heidelberg.de}  
}  
\begin{abstract}  
We give a status report of our current theoretical work on QCD near  
the light-cone.  
\end{abstract}  
  
\maketitle  
  
%%%%%%%%%%%%%%%%%%%%%%%%%%%%%%%%%%%%%%%%%%%%  
%% MAINMATTER  
%%%%%%%%%%%%%%%%%%%%%%%%%%%%%%%%%%%%%%%%%%%%  
  
\section{Introduction}  
  
A simple extension of the quantum mechanics of bound states to a  
relativistic field theory of massless quanta bound into hadrons is  
not possible. The light-cone Hamiltonian approach attacks this  
problem from a quite different point of view.   
Take a cube of length 2 fm filled with quarks and gluons   and  
boost it in the $3-$ direction  
with a Lorentz factor of $\gamma=1000$. This gedanken experiment is
well suited  
to imagine a proton moving with fast speed in the laboratory.   
The box will contract on one side, valence quark   
momenta will be high, and valence states will have very high energies.  
Naively vacuum properties of QCD are not important because of the  
high energies.  
By some suitable kinematic choices of  
coordinates one can construct invariants. Commonly, the light-cone  
energy $P^-=\frac{E-P_z}{\sqrt{2}}$ and the light-cone momentum    
$P^+=\frac{E+P_z}{\sqrt{2}}$ are chosen and $M^2= 2P_+P_--P_{\bot}^2$   
is invariant.   
With these variables all light-cone energies are positive  
and increase as $P^-=\frac {P_{\bot}^2+m^2 }{\sqrt{2} P^+}$ for small  
light-cone    
momenta.  
Only fluctuations with  small $P^+$  momenta may pose a problem. Their   
light-cone energies are very high. In light-cone physics the ultraviolet   
problem gets   
mixed up with the  infrared problem. Formally, the problem   
reappears in the context of constraint equations for $x^-$  
independent fields \cite{Franke:2004fr}.  
These constraint equations arise in the light-cone  
Hamiltonian framework, since the Lagrangian contains the velocities in  
linear form $L= \partial_- \phi \partial_+ \phi$. The momenta  
related to these velocities obey constraint equations including   
$\partial_- \phi$. Therefore, integrals of the equations of motion    
over the spatial light-cone distance $x^-$ become  
operator equations of reduced dimensionality (two transverse spatial  
dimensions and one time dimension). These equations are called  
zero-mode equations. For example, in equal time theory zero-mode equations  
determine the condensate of a scalar field.   
The $x^-$ independent zero-mode field couples to     
the transverse fluctuations of all other fields,   
consequently these equations depend on the cutoff  and are   
involved in the whole  
renormalization procedure. This feature is often overlooked  
in na\"{\i}ve pictures assuming either superrenormalizable models or  
models with a simple cutoff.   
In Nambu-Jona-Lasinio models one has been able to  
solve \cite{Itakura:2001yt,Lenz:2004tw}   
these zero-mode equations e.g. in large $N_c$ approximation   
giving a view of chiral symmetry breaking  
on the light-cone, which is quite special.  
These zero-mode equations have not been solved in QCD.

A common argument goes as follows: Zero modes   
decouple from the rest of the theory, because  their energies lie  
beyond the cutoff. Na\"{\i}vely, the light-cone momentum $P^+=0$ means   
that the light-cone energy $P^-=\infty$.  
If, however, the mass $m$ of the zero mode is zero, the  
mode does not disappear into infinity for very small transverse momenta.  
How is the situation in QCD? Can we just ignore this problem,  
buy a big computer, use some suitable Fock truncation, put  
all transverse gluon modes into a Hamiltonian matrix and  
diagonalize it?   
Pauli and Brodsky \cite{Hornbostel:1988fb} and many  
others have solved successfully 1+1 dimensional theories.      
QCD on the light-cone is a tremendously seductive field theory, since  
the Euler-Lagrange equation for time-like   
light-cone potential can be solved directly   
in a  gauge, where the potential  along the spatial light-cone  
direction vanishes. The resulting Hamiltonian contains the light-cone  
Coulomb energy  plus the kinetic energies of the   
transverse gluons and nothing else. The light-cone Coulomb  
energy is already in a form which linearly confines sources separated along  
the spatial light-cone directions. This is a simple consequence of  
the massless gluon propagator in one spatial dimension.

The massless gluon interaction has to be  
implemented also correctly for colored line sources smeared over   
the spatial light-cone direction. Otherwise, we violate the  
equal treatment of all spatial directions. This necessity can be  
demonstrated rather easily in perturbation theory, where the  
rotational invariance of the gluon exchange is reconstituted via the  
exchange of one transverse gluon.   
I think, one can be easily misled by the   
experience that QCD will always favor a finite correlation mass   
for color sources moving along time-like directions. At finite  
energies one sees this phenomenon in the hadronic cross sections     
which are given by the geometrical sizes of the hadrons, the  
low light-cone momentum partons do not matter at finite (small)  
energies.   
There is a natural  
transverse scale of the moving proton. The energy   
dependence of the high $Q^2$-structure functions indicate, however,  
an abnormal increase of ``size'' in transverse direction. The proton   
first gets blacker, but then its transverse radius has to   
increase.   
Purely theoretical arguments point towards conformal  
invariance at high energies, a conjecture,  which supports  
the view, that  
partons with small light-cone momenta sampling large spatial  
light-cone distances    
correlate over large transverse distances compared to  
normal hadronic scales.

We have analyzed QCD    
approaching the light-cone with a tilted near-light-cone coordinate  
reference system \cite{Ilgenfritz:2000bj,Naus:1997zg,Pirner:2002fe}   
containing a parameter   
$\eta \neq 0$ giving the distance away  
from the light-cone. The constraint equations appear in the near-light-cone  
Hamiltonian as terms proportional to $1/\eta^2$.  
We then multiply the light-cone energy with $\eta$,  
considering $\tilde P_+=\eta T^+_+$ and divide   
the light-cone momentum by $\eta$, defining  
$\tilde P_-=\frac{1}{\eta} T^+_-$. The invariant masses   
remain unchanged  
up to terms higher order in $\eta$. 
By the trick with near-light-like  
coordinates we can derive   
a full quantum Hamiltonian for the zero modes which now   
depends on the QCD coupling  $g$, the extension $L_{||}$ of the   
spatial light-cone distance compared to some lattice size $a$ (or  
ultraviolet cutoff $\Lambda=1/a$) and the parameter $\eta$ which gives the  
nearness to the light-cone.  
Having fixed the QCD coupling $g$ which determines the lattice size  
$a$, we would like to study in this Hamiltonian the physics at large  
longitudinal distances $L_{||}/a \rightarrow \infty$   
close to  the light-cone $\eta\rightarrow 0$.   
Because of dimensional reduction the product   
\begin{equation}  
s=\frac{\eta L_{||}}{a}  
\end{equation}  
appears as a coupling in the Hamiltonian.  
Its limit is not defined. The order of  
the limiting process is important as one knows from  
simple superrenormalizable models.   
One first has to let $L_{||}/a \rightarrow \infty$  
and then   $\eta\rightarrow 0$ in order not to lose the nonperturbative  
properties of the vacuum. For QCD an analytical limiting process is  
impossible. Therefore, the only way out is to start for large $s$,  
corresponding to fixed $\eta$ and  large $L_{||}$ and then approach  
smaller values of $s$.

This procedure ends, when we have found a  
fixed point  $s^*=\frac{\eta L_{||}}{a}$ where    
the mass gap of the zero mode theory vanishes.   
Approaching this fixed point from the  
correct side which corresponds to a  large longitudinal  
extension of the lattice, we include  the nonperturbative dynamics of  
the zero modes. The trivial, wrong other side  
where $s$ is arbitrarily small would be disconnected   
from the large $L_{||}$ limit.  
When the (2+1)-dimensional   
system has an infinite correlation length,  
both the infrared limit of large longitudinal distances and  
of nearness to the light-cone is realized.  
For a simplified zero mode theory in SU(2) we have   
demonstrated such a possibility on the lattice  
\cite{Ilgenfritz:2000bj}.  
In principle the full (3+1)-dimensional theory can be solved   
for any $\eta$ as long   
$g,L_{||}/a$ are chosen in such a way that we have asymptotic  
scaling. But in  order to synchronize the infrared   
behavior encoded in the zero mode system correctly   
with the ultraviolet behavior of small lattice size, the choice of $\eta$  
is no longer free for a given length of the longitudinal  
direction, one must choose $\eta$ in agreement with the fixed  
point found in the zero mode calculation, i.e. in the   
(3+1)-dimensional calculation the number of slices $L_{||}/a$ in  
spatial light-cone direction determines $\eta$  
\begin{equation}  
\eta= \frac{s^*}{L_{||}/a}.  
\end{equation}

It has to be demonstrated numerically that with decreasing   
QCD-coupling $g$ the value $s^*$ becomes smaller in such a   
way that we approach the light-cone  $\eta\rightarrow 0$ having    
a reasonable number of slices  $L_{||}/a $   
in spatial light-cone direction.  
The reduced calculation in SU(2) \cite{Ilgenfritz:2000bj}   
  was done without the inclusion of  
transverse gluons, so we still have to prove that this procedure   
works.   
Phenomenologically \cite{Pirner:2002fe} we have conjectured that the  
increase of the high-energy electron-proton cross section is  
due to this critical point $s^*$.   
At infinite energies when this point  is approached,  
the correlation length of   
near-light-like Wilson lines of the partons increases with a  
critical index from $Z(3)$ symmetry. The photon density   
remains power behaved beyond the short distance scale given  
by the resolution of the photon. According to our conjecture   
this critical opalescence  
phenomenon is the cause of the increase of the virtual photon  
cross section with high energies.

\section{The QCD Hamiltonian and momentum}  
We use the near light-cone coordinates defined in ref. \cite{Naus:1997zg}

\BA  
x^+&=&\frac{1}{\sqrt2}((1+\frac{\eta^2}{2})x^0+(1-\frac{\eta^2}{2})x^3)  
\nonumber\\  
x^-&=&\frac{1}{\sqrt2}(x^0-x^3)   
\label{eq.(1)}  
\EA  
with the metric  
  
\BE  
g_{\mu\nu}=\begin{array}{cccc}  
0 & 0 & 0 & 1\\  
0 & -1 & 0 & 0\\  
0 & 0 & -1 & 0\\  
1 & 0 & 0 & -\eta^2 \end{array},  
g^{\mu\nu}=\begin{array}{cccc}  
\eta^2 & 0 & 0 & 1\\  
0 & -1 & 0 & 0\\  
0 & 0 & -1 & 0\\  
1 & 0 & 0 & 0 \end{array}  
\label{eq.(2)}  
\EE  
with $\mu,\nu=+,1,2,-,\det g= - 1$, which defines the scalar product  
  
\BA  
x_\mu y^\mu & = & x^-y^+ + x^+y^- - \eta^2 x^-y^- - \vec x_\perp\vec  
y_\perp\nonumber\\   
            & = & x_-y_+ + x_+y_- + \eta^2 x_+y_+ - \vec x_\perp\vec  
y_\perp.  
\label{eq.(3)}  
\EA  
We refer to $p_+$ as ``light-cone'' energy and to $p_-$  
as ``light-cone'' momentum. We restrict ourselves to the color gauge  
group $SU(2)$ and to gluons. The Lagrangian density in the near  
light-cone coordinates reads:  
  
\BA  
\cal L &=& \frac{1}{2}F^a_{+-}F^a_{+-}+\sum_{i=1,2}  
           (F^a_{+i}F^a_{-i}+\frac{\eta^2}{2}F^a_{+i}F^a_{+i})\nonumber\\  
       &&  -\frac{1}{2}F^a_{12}F^a_{12}.  
\label{eq.(4)}  
\EA  
  
The energy momentum tensor has the general form:

\BE  
T^{\mu\nu}=\sum_{a}(g^{\mu \alpha}g^{\rho \beta} F^a_{ \alpha \beta}
F^a_{\rho \nu}  
-\delta^\mu_nu g^{\rho \alpha} g^{\sigma\beta} F^a_{ \alpha \beta}  
F^a_{\sigma \rho})  
\EE  
  
We introduce dimensionless gauge fields and coordinates  
  
\BA  
\tilde A^a_\pm &=& ga_\parallel A^a_\pm, \qquad   
                             \tilde x_\pm=x_\pm/a_\parallel;\nonumber\\  
\tilde A^a_i   &=& ga_\perp A^a_i, \qquad  
                             \tilde x_\perp=x_\perp/a_\perp.  
\label{eq.(5)}  
\EA  
The dimensionless Lagrange density $\cal L$ has the form  
\BE  
\int {\cal L} d^4x=\int {\tilde{\cal L} }d^4\tilde x  
\EE  
with  
\BA  
\tilde{\cal L} &=&  
\frac{1}{2}\frac{1}{g^2}(\frac{a_\perp}{a_\parallel})^2  
\tilde F^a_{+-}\tilde F^a_{+-}\nonumber\\  
&& +\frac{1}{g^2}\sum_i(\tilde F_{+i}\tilde F_{-i}  
+\frac{\eta^2}{2}\tilde F_{+i}\tilde F_{+i})\nonumber\\  
&& -\frac{1}{2}\frac{1}{g^2}(\frac{a_\parallel}{a_\perp})^2  
\tilde F^a_{12}\tilde F^a_{12}.  
\label{eq.(6)}  
\EA  
The field momenta conjugate to $\tilde A_i$ and $\tilde A_{-}$ are  
\BA  
\tilde\Pi^a_i &=& \frac{\partial\tilde{\cal L}}{\partial\tilde F_{+i}}=  
\frac{1}{g^2}( \tilde F^a_{-i}+\eta^2\tilde F^a_{+i})\nonumber\\  
\tilde\Pi^a_- &=& \frac{\partial\cal L}{\partial\tilde F_{+-}}=  
\frac{1}{g^2}(\frac{a_\perp}{a_\parallel})^2\tilde F_{+-}.  
\label{eq.(7)}  
\EA  
>From now on we drop the tilde symbol from all coordinates, fields and  
momenta to facilitate the writing and reading of all formulas.  
The commutation relations between fields and momenta are standard. 
\BA  
&& [\Pi^a_i(\tilde x_-,  x_\perp,  x_+), A^b_j(   
y_-,  y_\perp,  x_+)]=\nonumber\\  
&& i\delta^{ab}\delta_{ij}\delta(  x_- -  y_-)\delta(   
x_\perp-  y_\perp)\nonumber\\  
&&[\Pi^a_-(  x_-,  x_\perp,  x_+), A^b_-(  y_-,   
y_\perp,y_+)]=\nonumber\\  
&&i\delta^{ab}\delta(  x_- -  y_-)\delta(  x_\perp-   
y_\perp).  
\label{eq.(8)}  
\EA  
The dimensionless light-cone energy density $T^+_{+}$ and  
light-cone momentum density $T^+_-$ are obtained from the   
energy momentum tensor $T^{\mu\nu}$ and the skewed metric

\BA  
P_+=\int  T^+_{+}d  x^-d  x_\perp \\  
P_-=\int  T^+_{-}d  x^-d  x_\perp  
\EA  
with  
\BA  
T^+_{+}&=&\frac{1}{2}  
g^2\left(\frac{a_\parallel}{a_\bot}\right)^2 \Pi^2_-\nonumber\nonumber\\  
&&+\frac{1}{2}\frac{1}{g^2}\left(\frac{a_\parallel}{a_\bot}\right)^2  
  F_{12}  F_{12}\nonumber\\  
&&+\frac{1}{2}g^2\frac{1}{\eta^2}( \Pi_i-\frac{1}{g^2}   
F_{-i})^2\nonumber\\  
T^+_{-}&=&\frac{1}{2}(\Pi_i F_{-i}+F_{-i}\Pi_i).  
\label{eq.(10)}  
\EA

The light cone energy and momentum   have an obvious  
symmetry. They have electric-magnetic duality of the transverse  
fields which any solution to the problem must  
respect.  
  
\BA  
\Pi_i&\to&\frac{1}{g^2}F_{-i}\nonumber\\  
\frac{1}{g^2}F_{-i}&\to&\Pi_i.  
\label{eq.(11)}  
\EA  
  
Furthermore, since in the Lagrangian $ {\cal L}$ there are no  
terms with time derivatives of $  A_+$, the field $  A^a_+$  
acts like a Lagrange multiplier for $G^a$, the Gauss law. For any  
wavefunction $|\Phi>$ of the system the following identity must hold.  
  
\BA  
G^a|\Phi>&=&(\frac{1}{g^2}\left(\frac{a_\bot}{a_\parallel}\right)^2   
D^a_-  F_{+-}^a\nonumber\\  
&&+\frac{1}{g^2}\sum_i  D_i\left(  F_{-i}^a+\eta^2   
F_{+i}^a\right))|\Phi>\nonumber\\  
&=& (  D^a_- \Pi^a_- +\sum_i D_{i} \Pi_i^a)|\Phi>=0.  
\label{eq.(12)}  
\EA  
  
This Gauss law is fulfilled as long as  only  
closed loops exist in the wave function, or in the case of excited links the  
electric flux must be conserved at each site, i.e. there are also  
multiple connected flux loops possible.   
  
If one chooses $a_\parallel << a_\bot$ and uses the same number of  
sites in  
$  x_-$ and $  x_\bot$ directions, one ends up with a real system,  
which is contracted in the longitudinal directions. Verlinde and  
Verlinde \cite{Verlinde:1993te}, and Arefeva \cite{Arefeva:1993hi},   
have advocated such a set up to describe high  
energy scattering. A contracted lattice means the minimal momenta  
become high in longitudinal direction and this looks a promising  
starting point for high energy scattering.  
  
One sees from the Lagrangian $ {\cal L}$  in eq. (\ref{eq.(6)}) that  
the limit $a_\parallel/a_\bot\to 0$ enhances the terms with   
$  F_{+-}  F_{+-}$ and suppresses transverse  
$  F_{12}  F_{12}$.  
  
Because of the enhanced  couplings Verlinde and Verlinde conclude that the  
curvature in longitudinal directions is zero. One ends up with only  
one term which in the Hamiltonian is the term $\propto\frac{1}{\eta^2}$  
and fixes the dual symmetry of the electric and magnetic fields.

Our current project \cite{zeuthen} is to introduce lattice variables
 into this  
framework and solve the $1/\eta^2$ part of the Hamiltonian exactly.

\section{A valence-quark light-cone Hamiltonian}

In this section, I  would like to present a   
derivation where the near-light-cone method and the    
field strength correlators work    
nicely together. This example    
demonstrates their practicality as a calculational and heuristic tool.  
Firstly, one can analytically do the calculation in the stochastic vacuum  
model and secondly, the result is so close to reality that one  
can see the model-independent result.   
In our application of the stochastic vacuum model  
to high-energy scattering we always use Wilson loops  
which are on the light-cone. The expectation values of a Wilson loop along  
the light-cone is unity, because the area of a light-like   
Wilson loop is zero. I was always disturbed by this  
fact, because I thought that a color dipole moving with  
the speed of light should  feel  confining forces.    
The wavefunction renormalization due to single loops   
cancels out in the $S$-matrix, but the puzzle remained  
to me. So recently, Nurpeissov and myself \cite{Pirner:2004qd} have   
looked into this problem again using a tilted Wilson loop   
corresponding to a fast moving dipole  in Euclidean  
and in Minkowski space, i.e. we applied the near light-cone trick.

In Euclidean space the Wegner-Wilson loop can be represented with the help of   
the Casimir operator in the  
fundamental representation $C_2(3)=t^2=4/3$  
\begin{equation}  
\left<W[C]\right>_G\,=\,\exp{\left[-\frac{C_2(3)}{2}\chi_{ss}\right]}.  
\label{eq_15}  
\end{equation}  
We  calculate  $\chi_{ss}$ as the double area integral of the correlation  
function over the surface of the loop.   
Let us consider the $\chi_{ss}$ function  for large separations   
$R_0$ of the quark and antiquark,   
where the confinement term plays the main role.   
For the nonperturbative (NP) confining (c)   
component $\chi^{NP\,c}_{ss}$ we get the   
following expression for large  distances $R_0 \alpha>> 2 a $  
\begin{equation}  
\chi^{NP\,c}_{ss}= \lim\limits_{T\rightarrow\infty}{}  
        \left[\frac{2\pi^3 a^2 G_2 \kappa T}{3(N_c^2-1)}\cdot R_0\alpha  
        \right].  
\end{equation}  
Here $G_2$ denotes the gluon condensate, $\kappa$ the weight of the  
confining correlator compared to the nonconfining correlator, $a$ gives  
the correlation length and $T$ the extension of the loop in Euclidean time.  
%The orientation of the loop is shown in     
%(Fig. \ref{loop_eu}).  

The geometry of the arrangement enters into the factor $\alpha$.
The angle $\theta$ gives the tilting of the loop in the $X_3,X_4$
plane. The angle $\phi$ defines the angle of the $q\bar q$ connection
in the $X_1,X_3$ plane.

\begin{equation}  
\alpha=\sqrt{1-\cos^2{\phi}\sin^2{\theta}}.  
\end{equation}

One recognizes that the confining interaction leads to a VEV of the  
tilted  
Wilson loop which is consistent with the area law for large distances  
$R_0$  
\begin{eqnarray}  
<W[C]>&=&e^{-\sigma R_0 \alpha T}\\  
\sigma&=&\frac{\pi^3 G_2a^2 \kappa}{18},  
\end{eqnarray}  
where $\sigma $ is the string tension   and the  
area is obtained from  
\begin{eqnarray}  
Area&=& T R_0 \int_{-1/2}^{1/2}du \int_0^1 dv   
\sqrt{\left(\frac{dX_{\mu}}{du}\right)^2\left(\frac{dX_{\mu}}{dv}\right)^2  
-\left(\frac{dX_{\mu}}{du}\frac{dX_{\mu}}{dv}\right)^2}\\  
    &=&T R_0 \alpha.  
\end{eqnarray}  
For the  
Wegner-Wilson loop in Minkowski space-time we define $\chi_{ss}$  in  
the following  way    
\begin{equation} \left<W[C]\right>_G =  
\exp{\left[-i\,\frac{C_2(3)}{2}\chi_{ss}\right]}.    
\end{equation}   
Minkowskian geometry enters via the factor  
\begin{equation}  
\alpha_M=\sqrt{1+\cos^2{\phi}\sinh^2{\psi}},  
\end{equation}  
which is consistent with the  analytical continuation of  
the Euclidean  expression  
$\alpha=1-\cos^2{\phi}\sin^2{\theta}$ into Minkowski space  
by transforming the angle $\theta \rightarrow i \psi$.  
This analytical continuation is similar to the analytical  
continuation used in high-energy scattering   
\cite{Meggiolaro:1996hf,Meggiolaro:2001xk,Hebecker:1999pb},  
where the angle between two Wilson loops transforms in   
the same way.

The confining contribution to $\chi_{ss}$ reads in Minkowski space:  
\begin{eqnarray}  
\chi^{NP\,c}_{ss}= \lim\limits_{T\rightarrow\infty}{}  
        \left[\frac{2\pi^3 a^2 G_2 \kappa T}{3(N_c^2-1)}\cdot R_0\alpha_M  
        \right].  
\end{eqnarray}  
In order to interpret this result, one must  
define  
the four velocities of the particles described by the tilted loop  
\begin{equation}  
u_{\mu}= (\gamma,0_\bot,\gamma \beta).  
\end{equation}  
The exponent giving the expectation value of the Wilson loop acquires a new  
meaning now, since $- ig \int d\tau A^{\mu}u_{\mu} =- ig \int d\tau   
( \gamma A^0- \gamma \beta A^3)$,  
which leads in the VEV to a value for $ \beta \approx 1$  
\begin{equation}  
<W_r[C]>=e^{-i \gamma(P^0-P^3) T}.  
\end{equation}  
  
The light-cone energy arising from the confining part of the  
correlation function has the form  
\begin{equation}  
P^-= \frac{1}{\sqrt{2}}\left(\sigma R_0  
\sqrt{\cos(\phi)^2+\sin(\phi)^2/  
\gamma^2}\right).  
\end{equation}  
One sees that the Wilson loop for boosts with large $\gamma$ indicates  
that the light-cone energy  
does not  depend on the transverse distance $R_0 sin \phi$    
between the quarks.  
We introduce the relative $+$ momentum $k^+$ and transverse  
momentum $k_{\bot}$ for the quarks with   
mass $\mu$.   
By adding the above ``potential'' term to the kinetic term of relative motion   
of the  two particles we complete the Hamiltonian $P^-$  
\begin{equation}  
P^-= \frac{(\mu^2+ k_{\bot}^2)P}{2 (1/4 P^2-k^{+2})}+  
\frac{1}{\sqrt{2}}\sigma \sqrt{x_3^2+x_{\bot}^2/\gamma^2}.  
\end{equation}  
Next, we multiply $P^-$ with the plus component of the momentum $P^+$ and use that  
$P^+/M=\sqrt {2} \gamma M$ to eliminate the boost variable from the  
Hamiltonian.  
Further, we introduce   
the fraction  
$\xi=k^+/P^+$ with $|\xi| < 1/2$ and   
its conjugate  the scaled longitudinal space  
coordinate $\sqrt{2} \rho= P^+ x_3$ as dynamical variables.  
For our configuration the relative time of the quark and antiquark is zero.  
Then we get the light-cone Hamiltonian in  
a Lorentz invariant manner, because the variables $\xi,\rho,k_{\bot}$  
and  $x_{\bot}$ are invariant under boosts  
\begin{equation}  
M^2=2 P^+ P^-=  \frac {(\mu^2+ k_{\bot}^2)} {1/4-\xi^2} +   
2 \sigma\sqrt{\rho^2+ M^2 x_{\bot}^2}.  
\end{equation}  
To solve  
the $M^2$ operator one has to replace the square root  
operator by introducing an auxiliary parameter $s$ of dimension mass squared  
and minimize $M^2$ with respect to  
variations of $s$. Final self consistency must be reached with a guessed mass eigenvalue $M_0$  
\begin{equation}   
M^2=  \frac {(\mu^2+ k_{\bot}^2)} {1/4-\xi^2} +   
\frac{1}{2}\left( 4 \sigma^2 \frac {\rho^2+ M_0^2 x_{\bot}^2}{s} +s \right).  
\end{equation}  
In addition, one has to put the self-energy correction calculated  
by Simonov \cite{Simonov:2001iv}   
which is $\Delta \mu^2= -4 \sigma *f(m_q)/\pi$ and get  
\begin{equation}  
M^2=  \frac {(\mu^2 - \frac{4 \sigma f}{\pi}+ k_{\bot}^2)} {1/4-\xi^2} +   
\frac{1}{2}\left( 4 \sigma^2 \frac {\rho^2+ M_0^2 x_{\bot}^2}{s} +s \right).  
\end{equation}  
For light quarks the function $f(m_q)$ is close to unity. We have used  
the above equation with a simple trial function:  
\begin{equation}  
\psi( \xi, x_{\bot}) = N cos( \xi \pi)e^{-\frac{x_{\bot}^2}{2 x_0^2}}.  
\end{equation}  
We   obtain two solutions \cite{schlaudt} with positive masses  
due to the $s$-minimization. One solution is  
very low in mass and the other rather high.  
By tuning $f(mq)=0.8615$ away from unity the lower solution is pion-like with  
a really low mass, whereas the other solution lies at a   
typical hadronic scale  
\begin{equation}  
M_{low}=0.138 GeV\\  
\end{equation}  
\begin{equation}  
M_{high}=1.1 GeV\\ .  
\end{equation}  
Since on the light-cone   
the mechanism of chiral symmetry breaking is of particular interest,  
we would like to understand this result better. In the approach  
given here confinement plays an important role in contrast to  
Nambu-Jona-Lasinio effective models, which give an adequate description  
of spontaneous chiral symmetry breaking but do not include  
confinement.    
  
The confining interaction in the light Hamiltonian was derived in the  
specific model of the stochastic vacuum. But it also can be inferred  
from the simple Lorentz transformation properties of the phase in the  
Wilson loop and a lattice determination of the tilted Wilson  
expectation values. In this respect the final Hamiltonian is model  
independent.  
  
The inclusion of confining forces in the initial and final state wave  
functions can put all scattering cross sections   
calculated with the stochastic vacuum model on a much safer base, when   
wave functions and cross sections are derived consistently.  
For low $Q^2$ photon wave functions the long-distance  
part of the wave function matters strongly and confinement is  
important cf. \cite{Dosch:1997nw}.  
Especially the diffractive cross section  
has a large contribution from large dipole sizes and a correct   
behavior can only be expected when the problem of the large  
dipole wave function is treated adequately.   
Another extension of the above calculation is the coupling of the   
initial $q \bar q$ state to higher Fock states $q \bar q g$ with  
gluons which can be calculated   
with Wilson loops near the light-cone in Minkowski space.

\vspace{1.0cm}          
\section{Discussion and Conclusions}

I have tried to give some impression how QCD appears near  
the light-cone. I think we have now a calculational framework to  
approach the light-cone in a systematic way. It does not look   
much easier than equal time lattice gauge theory. One may hope  
that some simplifications arise in the process of studying it.  
The work  on a Wilson loop near the light-cone looked very complicated  
and intransparent at the beginning, but it reduced to some   
simple form. I like this example because it shows how the  
vacuum acts near the light-cone.

%\markright{References}  

%\newpage  
%  
%\begin{figure}[ht]  
%\includegraphics[height=5.0\textheight]{fig_1}  
%\caption{Configuration of the Wegner-Wilson loop in Euclidean space-time.}  
%\label{loop_eu}  
%\end{figure}  

\end{document}